# Designing the Color of Hot-Dip Galvanized Steel Sheet Through Destructive Light Interference Using a Zn-Ti Liquid Metallic Bath

Gabor Lévai, Melinda Godzsák, Tamás I. Török, József Hakl, Viktor Takáts, Attila Csik, Kálmán Vad, George Kaptay

The color of hot-dip galvanized steel sheet was adjusted in a reproducible way using a liquid Zn-Ti metallic bath, air atmosphere, and controlling the bath temperature as the only experimental parameter. Coloring was found only for samples cooled in air and dipped into Ti-containing liquid Zn. For samples dipped into a 0.15 wt pct Ti-containing Zn bath, the color remained metallic (gray) below a 792 K (519 °C) bath temperature; it was yellow at 814 K ± 22 K (541 °C ± 22 °C), violet at 847 K ± 10 K (574 °C ± 10 °C), and blue at 873 K ± 15 K (600 °C ± 15 °C). With the increasing bath temperature, the thickness of the adhered Zn-Ti layer gradually decreased from 52 to 32 micrometers, while the thickness of the outer $TiO_2$ layer gradually increased from 24 to 69 nm. Due to small Al contamination of the Zn bath, a thin (around 2 nm) alumina-rich layer is found between the outer $TiO_2$ layer and the inner macroscopic Zn layer. It is proven that the color change was governed by the formation of thin outer $TiO_2$ layer; different colors appear depending on the thickness of this layer, mostly due to the destructive interference of visible light on this transparent nano-layer. A complex model was built to explain the results using known relationships of chemical thermodynamics, adhesion, heat flow, kinetics of chemical reactions, diffusion, and optics. The complex model was able to reproduce the observations and allowed making predictions on the color of the hot-dip galvanized steel sample, as a function of the following experimental parameters: temperature and Ti content of the Zn bath, oxygen content, pressure, temperature and flow rate of the cooling gas, dimensions of the steel sheet, velocity of dipping the steel sheet into the Zn-Ti bath, residence time of the steel sheet within the bath, and the velocity of its removal from the bath. These relationships will be valuable for planning further experiments and technologies on color hot-dip galvanization of steel by Zn-Ti alloys.

DOI: 10.1007/s11661-016-3545-0

## I. INTRODUCTION

HOT-DIP galvanizing is a well-established surface coating industrial process with many scientific aspects studied in the past and even today.[1–9] In addition to obvious corrosion requirements, the interest in decorative use of colored coatings has recently increased.[10–12] These two fields have been combined in some patents and research papers.[13–22] Following this line, the goal of the present paper is to prepare colored coatings on hot-dip galvanized steel sheet using a Zn-Ti bath. As will be proven, color appears due to interference of light on a thin (30 to 80 nm), transparent $TiO_2$ layer, to be formed when the steel sample is removed from the Zn-Ti bath to air. The novelty of this paper is that our experimental findings are supported by a new, complex model, which connects various experimental parameters with the color of the coating.

## II. MATERIALS AND EXPERIMENTAL DETAILS

Low-alloyed steel sheets of type DC01 (St12 according to DIN) were selected in this work (see Table I for chemical composition) to avoid possible interactions with the alloying elements of steel. The dimensions of steel sheets to be coated were 0.8 mm thickness × 80 mm width × 100 mm length (the samples were dipped into the Zn-Ti bath along their length).

Special High Grade (SHG) zinc was used in this work (see Table II for its chemical composition). One can see that all contaminants were kept at a very low level. In our work, the extremely low levels of aluminum and

GABOR LEVAI, formerly Ph.D. Student with the University of Miskolc, Egyetemvaros, Miskolc, 3515 Hungary, is now Research Engineer with Innocenter Nonprofit Ltd, 7 Egyetem ut, Miskolc, 3515 Hungary. MELINDA GODZSÁK, Ph.D. Student, and TAMAS I TÖRÖK, Professor, are with the University of Miskolc, Egyetemvaros. JOZSEF HAKL and ATTILA CSIK, Senior Researchers, VIKTOR TAKÁTS, Junior Researcher, and KALMAN VAD, Senior Researcher, Head of Department, are with the Institute for Nuclear Research (MTA ATOMKI), 18/c Bem tér, Debrecen, 4026 Hungary. GEORGE KAPTAY, Professor, is with the University of Miskolc, Egyetemvaros, and also with the Bay Zoltan Ltd, BAYLOGI, 2 Igloi ut, Miskolc, 3519 Hungary. Contact mail: kaptay@hotmail.com

magnesium were of special importance, as at a higher level, Al and/or Mg might be oxidized preferably instead of Ti, not allowing the formation of $TiO_2$ and its corresponding color on the outer surface of the Zn coating.[20]

To alloy liquid zinc with titanium, titanium shavings were used for their high specific surface area to allow fast dissolution into the zinc bath. The chemical composition (see Table III) of the Ti shavings was found by ICP spectrometry (after dissolving them in an aqueous solution), using a 720-ES type equipment of Varian Inc. The calibration solutions were prepared using high-purity titanium and standard solutions Certipur IV of Merck. The Ti shavings were found to contain 1.95 wt pct Si, 0.21 wt pct Al and 0.020 wt pct Mg (see Table III). However, in the final Zn-0.15 wt pct Ti alloy, the Al content was only 0.0013 wt pct, while the Mg-content was only 0.00015 wt pct, which turned out to be sufficiently low for achieving colors during hot-dip galvanizing of our steel samples.

The Zn-Ti alloys were prepared in a cylindrical SiC crucible (diameter 160 mm, height 200 mm). The crucible was heated by an Al 1870 type resistance furnace (Hőker Ltd). The Zn-Ti alloy was prepared at 873 K (600 °C) in air. A zinc charge of 2 kg in weight was melted in the SiC crucible, and Ti shavings were added to it.

Alloying liquid zinc by the titanium shavings was a challenge, as the melting point of Ti (1941 K = 1668 °C) is much higher than the temperature interval of stable liquid Zn at 1 bar pressure [692 K–1180 K (419 °C–907 °C)]. Furthermore, the density of Ti (4.54 g/cm$^3$ at room temperature[23]) is much lower than that of liquid Zn (6.58 g/cm$^3$ at its melting point[24]), and so if the Ti shavings were just put on the surface of liquid Zn, they would be floating and oxidized on its surface, leading to unsuccessful alloying. Therefore, the cleaned and weighted Ti shavings were added on the surface of the Zn melt and immediately pushed under the level of molten Zn by an alumina rod and mixed by this rod into the liquid zinc bath. After mixing the system for 20 minutes, a sample was taken for analysis by ICP spectroscopy (for its description see above) and the alloy was casted and cooled spontaneously. In this way, different Zn-Ti alloys were prepared with Ti contents varying between 0.05 wt pct and 0.35 wt pct. During alloying, some of added Ti was lost due to partial oxidation, as evidenced by ICP measurements and by the different colors appearing on the surface of the Zn-Ti bath (pure Zn bath does not provide color except metallic, even if oxidized).

Prior to hot-dip galvanizing the steel samples, their surface was cleaned to ensure its good wettability and adhesion by the Zn-Ti liquid alloy. First, the steel samples were degreased during 5 minutes by a Dexacid H420 solution. Secondly, they were pickled during 5 minutes in the 25 to 37 wt pct HCl aqueous solution to remove the oxide layer from its surface. Thirdly, they were rinsed by tap water. After drying, the surface was fluxed during 1 minute by Fluorflux SPG aqueous solution (Floridienne Chimie S.A) and dried. Then, the final drying and pre-heating of the sample was performed right above the hot Zn-Ti alloy, using its waste heat. After 5 minutes of this pre-heating operation, the steel sample was immersed into the Zn-Ti liquid alloy for hot-dip galvanization.

The liquid Zn-Ti alloy was kept in the same SiC crucible and heated by the same electric furnace as used for the preparation of the Zn-Ti alloy (see above). Bath temperature was controlled by the thermocouple immersed into the Zn-Ti bath and protected by a thin alumina cover. The steel sample was held by a special sample holder, moved up and down by an electric motor with a desired speed. As a slag layer covered the Zn-Ti liquid alloy, it was shifted aside to allow immersion and removal of the steel sample by a special, not surface-treated steel sheet (so the Zn-Ti alloy did not adhere to its oxidized surface). The surface of the liquid alloys was cleaned in this way only on one side of the steel sample right before removing it from the liquid Zn-Ti bath. The steel sample was kept vertically and cooled spontaneously in still air to room temperature. In this way, at least one side of the hot-dip galvanized steel sample was obtained slag free. All further evaluations and measurements were conducted on this slag-free side of the sample.

The color of the steel samples was examined by human eye in natural light. Additionally, the color was measured by a CIE-Lab Hunter system, using a Konica-Minolta CM-2600d equipment. The concentration depth profile of the elements and the thickness of the Zn-Ti coatings on the steel samples were measured by glow discharge optical emission spectroscopy (GD-OES), using a Horiba Jobin Yvon GD-profiler 2 equipment. The outer surface of the Zn-Ti coating was analyzed in details by a secondary neutral mass spectrometer (SNMS) of type INA-X (Specs GmbH).

Table I. Chemical Composition of the Major Elements in DC01 Steel (from EN 10130:2006)

| Element | C | P | S | Mn |
|---|---|---|---|---|
| Max weight pct | 0.12 | 0.045 | 0.045 | 0.60 |

Table II. Chemical Composition of the Major Elements in SHG Zinc

| Element | Pb | Cd | Fe | Cu | Sn | Al |
|---|---|---|---|---|---|---|
| Max weight pct | 0.003 | 0.003 | 0.002 | 0.001 | 0.001 | 0.001 |

Table III. Chemical Composition of the Major Elements in the Titanium Shavings

| Element | Si | Al | Fe | Mg | Cu | Others |
|---|---|---|---|---|---|---|
| Weight pct | 1.95 | 0.21 | 0.046 | 0.020 | 0.016 | below 0.015 |

In the series of preliminary experiments, the majority of experimental parameters were varied, and finally fixed as the Ti content of the alloy was 0.15 wt pct; the speed of immersion of the steel sample into the Zn-Ti bath was 10 mm/s; the holding time of the steel sample in the Zn-Ti bath was 30 seconds; the speed of removal of the steel sample from the Zn-Ti bath was 10 mm/s; and atmosphere was normal air at room temperature. These fixed parameters allowed perform the hot-dip galvanizing of 220 steel samples in a reproducible way. The only experimental parameter varied in those 220 experiments was the temperature of the Zn-Ti bath, selected in the interval of 703 K to 923 K (430 °C to 650 °C). Several parallel experiments were conducted at the same temperature to confirm the reproducibility of the color.

With each steel sheet galvanized in the large Zn-Ti bath, some small amount of this bath was removed. However, the composition of this adhered Zn-Ti layer was the same as that of the large bath. Therefore, only the amount of the initial Zn-Ti bath reduced somewhat during the course of different galvanizing experiments, but the composition of this bath remained approximately the same. This was proven by ICP analysis of the Zn-Ti bath before and after the series of galvanizing experiments.

## III. PRIMARY EXPERIMENTAL OBSERVATIONS

### A. The Color of the Samples as Seen by Human Eye

Under the conditions listed above, the color of the hot-dip galvanized steel samples remained metallic (silver/gray) below 782 K (519 °C) of bath temperature. The color was yellow at 814 K ± 22 K (541 °C ± 22 °C), violet at 847 K ± 10 K (574 °C ± 10 °C), and blue at 873 K ± 15 K (600 °C ± 15 °C) (see Figure 1 and Table IV). Above 888 K (615 °C) of bath temperature, the slag formation on the surface of the Zn-Ti bath was so fast and severe that samples of clean surface could not be produced. The same was true when the Ti content of the Zn bath was increased from 0.15 wt pct to 0.3 wt pct. These results were similar to those obtained by Le and Cui.[20]

Some experiments were performed using a Ti-free Zn bath, and some further experiments were performed in an inert, argon atmosphere. None of these experiments provided any color, being different from the metallic color of Zn. Thus, it was proven that the color was provided by the combination of the Ti content of the Zn bath and the oxygen content of the cooling gas. It follows that one of the titanium oxides should be responsible for the color of the hot-dip galvanized steel sample.

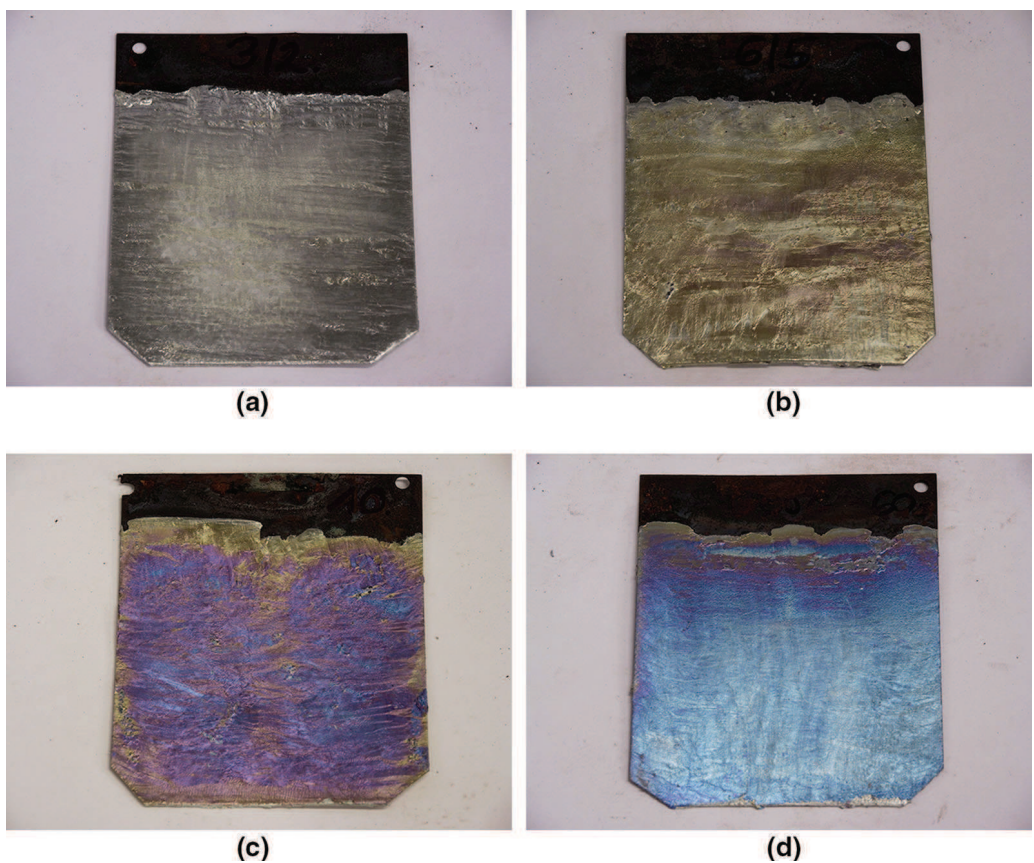

Fig. 1—Four steel sheets obtained from Zn-Ti baths of different temperatures: metallic (silver/gray) color is obtained at 723 K (450 °C) (*a*), yellow color is obtained at 813 K (540 °C) (*b*), violet color is obtained at 843 K (570 °C) (*c*), and blue color is obtained at 873 K (600 °C) (*d*).

Table IV. The Summary of Experimental Observations as a Function of Bath Temperature

| Method | Quantity/$T_{Zn}$ [K (°C)] | Below 792 (519) | 814 ± 22 (541 ± 22) | 847 ± 10 (574 ± 10) | 873 ± 15 (600 ± 15) |
|---|---|---|---|---|---|
| Visual | color | metallic | yellow | violet | blue |
| CIE-Lab | $\Delta b$ | — | 15 ± 7 | −9 ± 5 | −22 ± 7 |
| GD-OES | Zn-thickness ($\mu$m) | 52 ± 8 | 43 ± 7 | 38 ± 6 | 35 ± 6 |
| GD-OES | Ti on surface (wt pct) | 3 ± 2 | 14 ± 5 | 17 ± 5 | 35 ± 9 |
| SNMS | $TiO_2$ thickness (nm) | 24 ± 5 | 36 ± 7 | 54 ± 10 | 69 ± 14 |

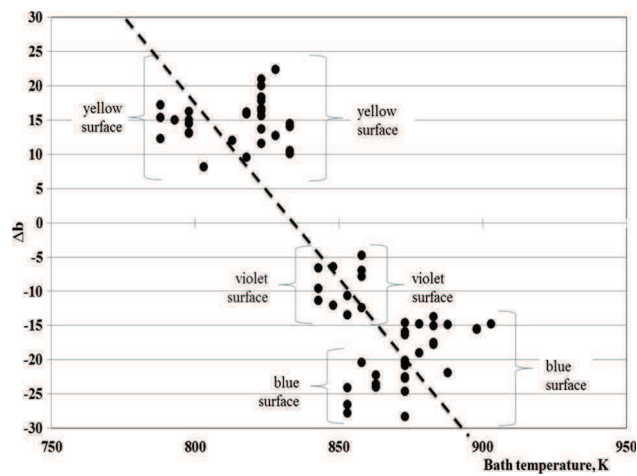

Fig. 2—The measured $\Delta b$ values on hot-dip galvanized steel samples as function of the bath temperature. Samples with yellow, violet, and blue colors (as seen by human eye) are shown in different, but slightly overlapping groups.

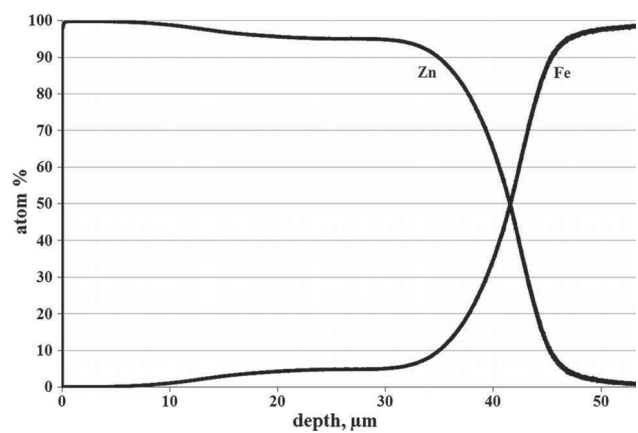

Fig. 3—Concentration profiles of different elements for a steel sample, dipped into a Ti-free Zn bath. The thickness of the Zn coating is measured at the interception of the curves for Zn and Fe. The concentration of other elements is below 1 at. pct.

### B. The Color Measured by a CIE-Lab Hunter System

As human eye uses three independent receptors to help our mind to create colors, color is measured also by three independent physical quantities by the CIE-Lab Hunter system. Among them, parameter b was selected, as this parameter describes the "color axis" between yellow and blue colors, observed in our samples. Parameter b was measured, respectively, to the metallic color; this relative value is called hereinafter parameter $\Delta b$. Yellow color is characterized by positive $\Delta b$ values, while blue color is characterized by negative $\Delta b$ values.

Table IV and Figure 2 show the measured $\Delta b$ values as function of the bath temperature. Samples with yellow, violet, and blue colors (as seen by human eye) are shown in different groups in Figure 2. One can see that a good linear correlation exists between the value of parameter $\Delta b$ and the bath temperature. This result confirms the results of visual observations.

### C. The Results of GD-OES Measurements

In Figure 3, the concentration profiles of chemical components are shown for a steel sample, dipped into a Ti-free Zn bath. The two major elements can only be seen in this figure, the iron and zinc, since the concentration of all other elements being present in the layer was <1 at.pct. The thickness of the coating was determined from the interception of the concentration curves. However, the "bumps" on the Zn- and Fe- concentration profiles between 10 and 30 $\mu$m depths indicate the formation of Zn-Fe intermetallic compound(s).

It is important to note that the thickness of the Zn layer is not a function of the vertical section of the sample. Also, no dropping of liquid Zn was observed during the cooling phase of the vertical steel sheet with liquid Zn layer around it. Thus, there is no significant flow of liquid Zn along the vertical steel sheet after it is removed from the Zn bath.

The full GD-OES concentration depth profiles for elements measured for a steel sample dipped into Zn-Ti baths and colored to violet are shown in Figure 4(a). It is important to note that there is no Ti peak found at the Zn/Fe interface. These spectra can be used to measure the thickness of the coating (see Table IV). However, in the enlarged spectrum of Figure 4(b), a new feature can be observed: a maximum in the Ti content in the outer 100 nm surface region. Its value measured for a violet sample was about 19 at.pct Ti. As follows from Table IV, the Ti content in the outer 100 nm of the coating increases with the increasing bath temperature. Thus, it is proven that the color change of the samples from metallic to blue (through yellow and violet) is connected with a gradual increase in the Ti content in the outer 100 nm of the Zn coating. Unfortunately, GD-OES measurements could not be properly calibrated for oxygen, so an oxygen peak cannot be seen in Figure 4(b). To correct this situation, a more accurate SNMS technique was applied (see next sub-chapter).

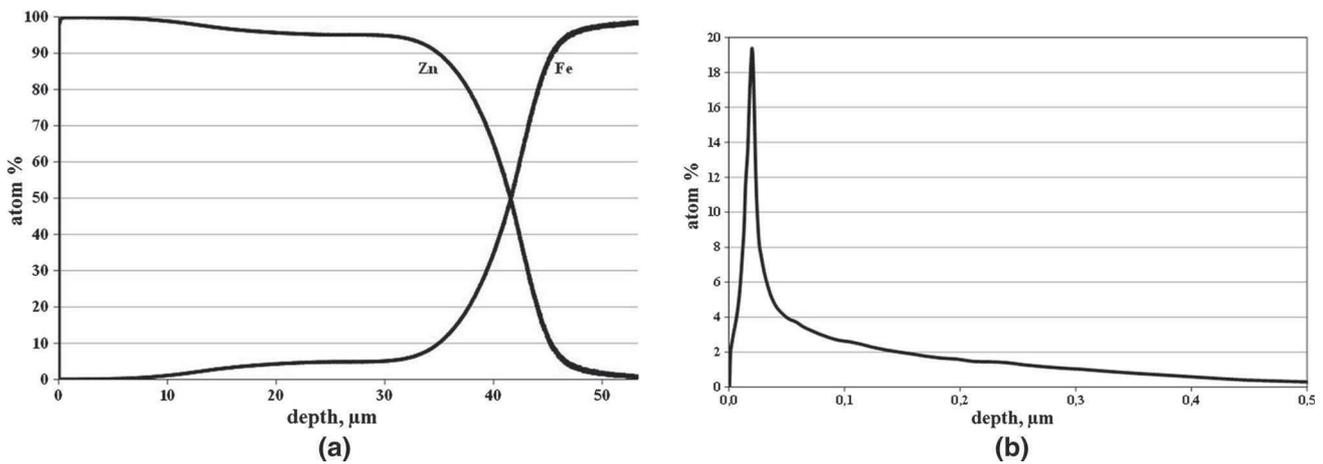

Fig. 4—Concentration profiles of different elements for a steel sample dipped into a Zn-Ti bath and colored to violet. The thickness of the Zn coating is measured at the interception of the curves for Zn and Fe in *a*. There is no Ti peak at the Zn/Fe interface. The concentration axis of *b* is enlarged by five times, showing a Ti peak in the upper 100 nm of the coating.

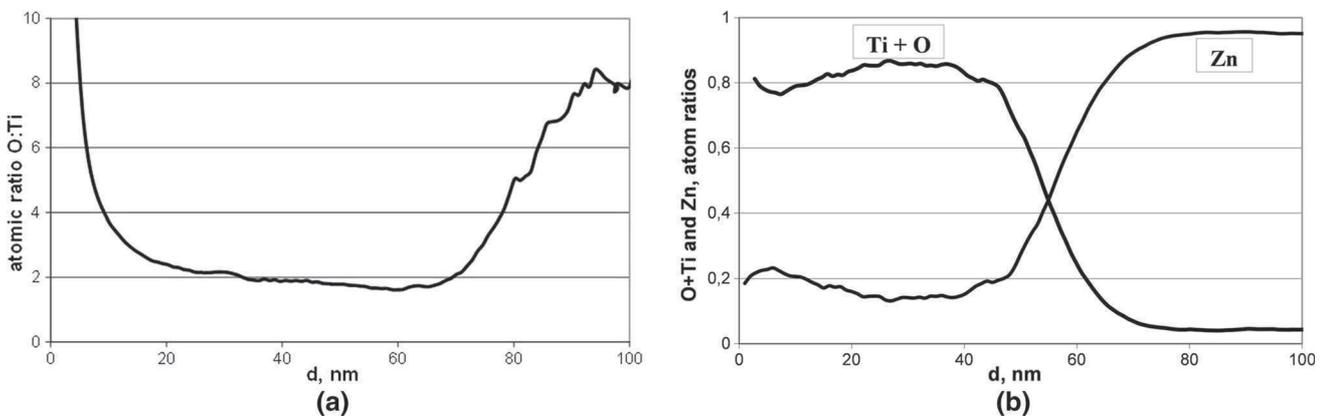

Fig. 5—Results of SNMS measurements performed on a blue sample. The atomic ratio of oxygen to titanium contents (*a*) and the concentration profiles of Zn and (Ti + O) (*b*) as function of the depth.

### D. The Results of SNMS Measurements

The upper colored surface of the sample was bombarded by $Ar^+$ ions. As a result of intermittent ion bombardment, the initially blue sample turned violet, and the initially violet sample turned yellow, while the initially yellow sample turned metallic. This experiment proved that the color of a sample originated from the outer coating and its thickness. Moving from a large thickness values toward smaller thickness values, the color changes in the following order: blue–violet–yellow–metallic. Combining this observation with the Ti content measured by GD-OES (see Table IV), one can conclude that the color is probably due to some Ti-compound of variable thickness.

Additionally, it was found by SNMS measurements that the outer region of the samples was mainly composed of three elements: zinc, oxygen, and titanium. The ratio of oxygen to titanium contents of a blue sample as a function of depth is shown in Figure 5(a). In the interval of depth between 20 and 70 nm, the O:Ti atomic ratio is around 2, proving the formation of $TiO_2$. This ratio became much higher in the topmost 20 nm thick layer, due to the surface roughness and surface contamination of the sample stored in air. The ratio becomes meaningless above 70 nm, as the outer $TiO_2$-rich layer was replaced by an inner, Zn-rich layer (see Figure 5(b)). In Figure 5(b), the concentration profiles of Zn and the sum of (O + Ti) are shown as measured by SNMS for a blue sample. To summarize, the outmost region of the Zn-Ti coating was rich in (Ti + O), while the inner region of the Zn-Ti coating was rich in Zn. The thickness of the $TiO_2$ coating estimated from Figure 5(b) was about $54 \pm 2$ nm. Repeating the same type of measurements on different samples, the thickness values of the outer $TiO_2$ layers are collected in Table IV for different bath temperatures. One can see that the thickness of the outer $TiO_2$ layer increases with the increase in the bath temperature.

In Figure 6, the measured Al signal from the upper 100 nm layer of the same blue sample is shown. A peak in Al content can be seen with its maximum at the depth of about $56 \pm 1$ nm. This Al enrichment is probably $Al_2O_3$, as Al has a stronger affinity for oxygen than titanium. From comparing it with the thickness of the TiO$_2$ layer (54 ± 2 nm from Figure 5(b)), two conclusions can be drawn: (i) the alumina layer is positioned between the outer TiO$_2$ layer and the inner Zn layer, and (ii) the thickness of the alumina layer is about 2 ± 1 nm.

### E. Summary of Experimental Observations

Coloring was found only when the sample was dipped into Ti-containing Zn bath and when it was cooled in air; coloring was not found in oxygen-free atmosphere and in Ti-free bath. Keeping certain experimental parameters constant (0.15 wt pct of Ti in the Zn bath and 10 mm/s removal speed of the steel sheet from the Zn-Ti bath to normal air, to mention just a few), the color remained metallic (gray) below 792 K (519 °C) of bath temperature; it was found yellow at 814 K ± 22 K (541 °C ± 22 °C), violet at 847 K ± 10 K (574 °C ± 10 °C), and blue at 873 K ± 15 K (600 °C ± 15 °C). With the increasing bath temperature, the thickness of the adhered Zn-Ti layer gradually decreased from 52 to 32 microns, while the average Ti content of the outer 100 nm layer increased from 3 to 35 wt pct. The outer surface of the Zn-Ti coating was proven to be TiO$_2$; with the increasing bath temperature, the thickness of this outer TiO$_2$ layer gradually increased from 24 to 69 nm. At the interface between the outer TiO$_2$ layer and the inner macroscopic Zn layer, a thin Al enrichment was found, which probably corresponds to the oxide Al$_2$O$_3$, with a thickness of about 2 nm.

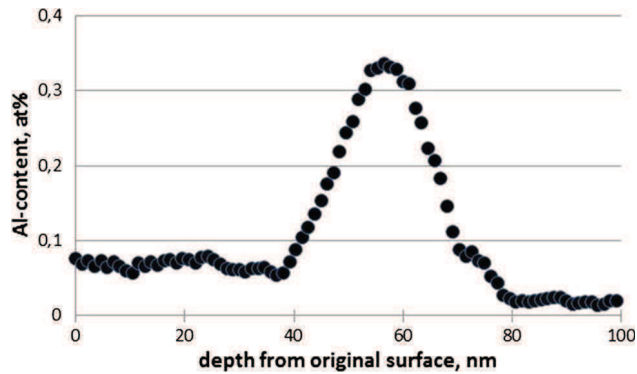

Fig. 6—Depth distribution of Al in the upper 100 nm of the coating (for the same blue sample as shown in Figs. 5(a) and (b).

Thus, the color was proven to be due to the formation of a thin TiO$_2$ layer. Different colors appear at different thicknesses of this layer, which were achieved by different bath temperatures in this work.

## IV. DISCUSSION

The model provided below will connect the reason (bath temperature increase) with the result (color change). There is obviously a long logical way from bath temperature to surface coloring. Therefore, a complex model is provided, using known facts of different branches of science: chemical thermodynamics, adhesion, heat flow, reaction kinetics, diffusion, and optics. The model predicts how the color changes as function of bath temperature and also other experimental parameters, not studied here. All model parameters, their symbols, units, and characteristic values used in our experiments are collected in Table V.

### A. Chemical Thermodynamics and Materials Balance

Basic thermodynamic properties of the Fe/Zn-Ti/O$_2$ system are collected in Appendix. First, let us find the minimum required partial pressure of oxygen needed to form a ZnO layer on the surface of pure liquid Zn, in accordance with the following chemical reaction:

$$Zn + 0.5 \times O_2 = ZnO. \quad [1a]$$

Reaction [1a] is accompanied by the following Gibbs energy change:

$$\Delta_{r1} G = \Delta_f G^o_{ZnO} - 0.5 \times R \times T \times \ln \frac{p_{O_2}}{p^o} - R \times T \times \ln a_{Zn}, \quad [1b]$$

where $p_{O_2}$ (bar) is the partial pressure of O$_2$ in the gas, $p_o$ = 1 bar is the standard pressure, $a_{Zn}$ (dimensionless) is the activity of Zn in pure liquid Zn, being 1 by definition. The condition of equilibrium is $\Delta_{r1} G = 0$. Substituting this condition into Eq. [1b], the required partial pressure of O$_2$ gas is as follows:

$$p_{O_2} = p^o \times \exp\left(\frac{2 \times \Delta_f G^o_{ZnO}}{R \times T}\right). \quad [1c]$$

Table V. Model Parameters and Their Characteristic Values

| Parameter | Symbol | Unit | Characteristic Value |
|---|---|---|---|
| Length of the steel sample | $L$ | m | 0.10 |
| Width of the steel sample | $w$ | m | 0.080 |
| Thickness of the steel sample | $d$ | m | $8.0 \times 10^{-4}$ |
| Ti content in the Zn bath | $C_{Ti}/x_{Ti}$ | wt pct / mole fraction | $0.15/2.0 \times 10^{-3}$ |
| Speed of sample removal from the bath | $v_{out}$ | m/s | 0.010 |
| Bath temperature | $T_{Zn}$ | K (°C) | 703–923 (430–650) |
| Air temperature above the bath | $T_{air}$ | K (°C) | 303 (30) |
| Forced air velocity above bath (air knife) | $v_{gas}$ | m/s | 0 |
| Oxygen content of air | $C_{O2}$ | vol pct | 21 |
| Pressure of air | $p$ | Pa | $10^5$ |

Table VI. Minimum Partial Pressure of $O_2$, Required for Reaction [1a], (Eq. [1c], Table AI)

| T [K (°C)] | 700 (427) | 800 (527) | 900 (627) | 1000 (727) |
|---|---|---|---|---|
| $p_{O_2}$, bar | $1.3 \times 10^{-42}$ | $6.0 \times 10^{-36}$ | $8.9 \times 10^{-31}$ | $1.2 \times 10^{-26}$ |

Table VII. The Minimum Required Ti Content (mole fraction) in Liquid Zn to Form Different Oxides on the Surface of the Liquid Zn-Ti Alloy in Air (Calculated by Eq. [2c] and Data of Tables AI and AII)

| T [K (°C)] | 700 (427) | 800 (527) | 900 (627) | 1000 (727) |
|---|---|---|---|---|
| $TiO_2$ | $3.9 \times 10^{-16}$ | $1.5 \times 10^{-14}$ | $2.5 \times 10^{-13}$ | $2.4 \times 10^{-12}$ |
| $Ti_4O_7$ | $1.1 \times 10^{-15}$ | $3.5 \times 10^{-14}$ | $5.3 \times 10^{-13}$ | $4.9 \times 10^{-12}$ |
| $Ti_3O_5$ | $2.1 \times 10^{-15}$ | $6.3 \times 10^{-14}$ | $8.8 \times 10^{-13}$ | $7.5 \times 10^{-12}$ |
| $Ti_2O_3$ | $5.5 \times 10^{-15}$ | $1.7 \times 10^{-13}$ | $2.4 \times 10^{-12}$ | $2.1 \times 10^{-11}$ |
| $TiO$ | $1.4 \times 10^{-11}$ | $2.0 \times 10^{-10}$ | $1.6 \times 10^{-9}$ | $8.6 \times 10^{-9}$ |

Calculations were performed by Eq. [1c], using the data of Table AI. The results are collected in Table VI. As follows from Table VI, the actual partial pressure of oxygen in normal air (0.21 bar) is much higher than the one required for Reaction [1a]. Thus, in the whole T interval of 700 K to 1000 K (427 °C to 727 °C), the surface of pure liquid Zn will be coated by a ZnO layer in air.

Now, let us suppose that some Ti is gradually added into the Zn bath. Let us calculate the minimum required Ti content, needed to form different Ti oxides instead of ZnO on the surface of the Zn-Ti alloy. For this, let us consider the following chemical reaction and its Gibbs energy change:

$$y \times ZnO + x \times Ti = Ti_xO_y + y \times Zn, \quad [2a]$$

$$\Delta_{r2}G = \Delta_f G^o_{Ti_xO_y} + y \times R \times T \times \ln a_{Zn} - y \times \Delta_f G^o_{ZnO} - x \times R \times T \times \ln(x_{Ti} \times \gamma^\infty_{Ti}), \quad [2b]$$

where x and y (positive integer numbers) are the stoichiometric coefficients of $Ti_xO_y$, $x_{Ti}$ (dimensionless) is the mole fraction of Ti in the Zn bath, and $\gamma^\infty_{Ti}$ is the activity coefficient of Ti (dimensionless) in its diluted solution of liquid Zn. The condition of equilibrium is $\Delta_{r2}G = 0$. Substituting this value into Eq. [2b] and taking into account $a_{Zn} \cong 1$, the required mole fraction can be found as

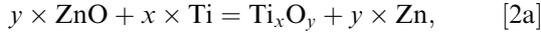

$$x_{Ti} = \frac{1}{\gamma^\infty_{Ti}} \times \exp\left(\frac{\Delta_f G^o_{Ti_xO_y} - y \times \Delta_f G^o_{ZnO}}{x \times R \times T}\right). \quad [2c]$$

Calculations are performed by Eq. [2c] using the data of Tables AI and AII (see Table VII). One can see that among all Ti oxides, the lowest Ti content in the Zn bath is needed to form $TiO_2$, so its formation has the highest probability. Moreover, the minimum Ti content required to form this oxide is lower by about 9 through 13 orders of magnitudes compared to the actual (relatively high) Ti content in our experiments (see Table V).

The growth of $TiO_2$ will be limited by the diffusion of ions through this layer (see below). The diffusion coefficients of different ions will be inversely proportional to their ionic radii,[30] which are equal[23] to 0.069 nm for $Ti^{+4}$, 0.083 nm for $Zn^{+2}$, and 0.132 nm for $O^{-2}$. Therefore, the diffusion coefficients of metallic ions are much higher than that of the oxygen ion. This means that the growth of the oxide layer is ensured by the diffusion of metallic ions through the oxide layer from the metallic phase toward the gas phase, i.e., the oxide layer thickens on its outside surface, as usual.[30] It is also clear that the diffusion coefficient of Ti ions through the oxide layer will be larger than that of the Zn ions, and so the outer part of the oxide layer will be $TiO_2$. However, at the inner $TiO_2$/Zn interface, the complex oxide can be formed (at least, in principle), according to the reaction:

$$2 \times TiO_2 + 2 \times Zn = Zn_2TiO_4 + Ti. \quad [3a]$$

The Gibbs energy change, accompanying Reaction [3a] is written as

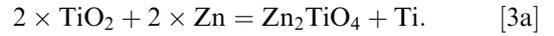

$$\Delta_{r3}G = \Delta_f G^o_{Zn_2TiO_4} + R \times T \times \ln(x_{Ti} \times \gamma^\infty_{Ti}) - 2 \times \Delta_f G^o_{TiO_2} - 2 \times R \times T \times \ln a_{Zn}. \quad [3b]$$

The condition of equilibrium is $\Delta_{r3}G = 0$. Substituting this value and $a_{Zn} \cong 1$ into Eq. [3b], the maximum mole fraction of Ti, allowing Reaction [3a] to take place, is found as

$$x_{Ti} = \frac{1}{\gamma^\infty_{Ti}} \times \exp\left(\frac{2 \times \Delta_f G^o_{TiO_2} - \Delta_f G^o_{Zn_2TiO_4}}{R \times T}\right). \quad [3c]$$

Calculations are performed with Eq. [3c] using the data of Tables AI and AII (see Table VIII). One can see that the actual Ti content in the Zn-Ti alloy (Table V) is larger by 8 to 12 orders of magnitude compared to the maximum Ti content allowing Reaction [3a] to happen (see Table VIII). Thus, the complex $Zn_2TiO_4$ will not be formed, as Zn is not able to replace Ti in its oxide. Finally, one can conclude that the Zn-Ti liquid alloy with at least $10^{-10}$ wt pct of Ti will be coated by a $TiO_2$ layer once exposed to air. This finding is in full agreement with the experimental results obtained by SNMS (see above).

Table VIII. The Minimum Required Ti Content (mole fraction) in Liquid Zn to Keep Stable $TiO_2$ Instead of $Zn_2TiO_4$ at the $TiO_2$/Zn Interface (Calculated by Eq. [3c] and Data of Tables AI and AII)

| T [K (°C)] | 700 (427) | 800 (527) | 900 (627) | 1000 (727) |
|---|---|---|---|---|
| $x_{Ti}$ | $2.6 \times 10^{-15}$ | $8.8 \times 10^{-14}$ | $1.4 \times 10^{-12}$ | $1.3 \times 10^{-11}$ |

Table IX. The Minimum Ti Content (mole fraction) in Liquid Zn, Required for Reaction [5a] (Calculated by Eq. [5c], Using Data of Tables AII and AIII)

| T [K (°C)] | 700 (427) | 800 (527) | 900 (627) | 1000 (727) |
|---|---|---|---|---|
| $x_{Ti}$ | 0.81 | 0.73 | 0.66 | 0.64 |

Now, let us suppose that all the Ti content of the Zn bath is diffused to its outer surface and formed $TiO_2$ with the oxygen of air, the latter being available without limit. Then, the maximum thickness of the oxide layer can be estimated from the material balance as

$$d_{TiO_2,max} = x_{Ti} \times \frac{V_{TiO_2}}{V_{Zn}} \times d_{Zn}, \quad [4a]$$

where $x_{Ti}$ (dimensionless) is the initial mole fraction of Ti in the Zn bath, $V_{TiO_2}$ = 19.0 cm$^3$/mol is the molar volume of $TiO_2$,[31] $V_{Zn}$ = 9.2 cm$^3$/mol is the molar volume of Zn,[23] and $d_{Zn}$ (m) is the thickness of the Zn-Ti layer adhered to the steel sample. Substituting the molar volume values into Eq. [4a],

$$d_{TiO_2,max} = 2.07 \times x_{Ti} \times d_{Zn}. \quad [4b]$$

If $d_{Zn}$ = 30 − 50 $\mu$m (see Table IV) and $x_{Ti}$ = 0.002 (see Table V), then from Eq. [4b], $d_{TiO_2}$, max = 130 − 210 nm. This would be the thickness of the $TiO_2$ layer, if all Ti from the Zn bath is consumed to form $TiO_2$. In reality, the measured values have the same magnitude but are considerably smaller: 20 to 80 nm (see Table IV). This is because the Zn coating solidifies quite fast when the steel sample is removed into cold air, and so there is no sufficient time to transform all Ti, dissolved in the Zn bath, into $TiO_2$.

Equation [4b] was derived supposing that Ti atoms are not stabilized at the Fe/Zn-Ti interface. However, the opposite is not excluded, as intermetallic compounds form in the Fe-Ti system.[26] Moreover, they are more stable than the intermetallic compounds in the Fe-Zn system (see Table AIII). Thus, let us consider the following simplified displacement reaction:

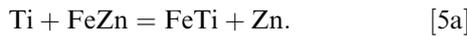

$$Ti + FeZn = FeTi + Zn. \quad [5a]$$

The Gibbs energy change of Reaction [5a] is written as

$$\Delta_{r5}G = \Delta_f G^o_{FeTi} + R \times T \times \ln a_{Zn} - \Delta_f G^o_{FeZn} - R \times T \times \ln(x_{Ti} \times \gamma^{\infty}_{Ti}). \quad [5b]$$

Reaction [5a] will be in equilibrium if $\Delta_{r5}G$ = 0. Substituting this condition into Eq. [5b] and taking into account $a_{Zn} \cong 1$, the minimum Ti content can be obtained to ensure Reaction [5a] as

$$x_{Ti} = \frac{1}{\gamma^{\infty}_{Ti}} \times \exp\left(\frac{\Delta_f G^o_{FeTi} - \Delta_f G^o_{FeZn}}{R \times T}\right). \quad [5c]$$

Calculations are performed by Eq. [5c] using the data of Tables AI and AII (see Table IX). One can see from comparison of Tables V and IX that the minimum Ti content required for the Reaction [5a] is higher by more than 2 orders of magnitude compared to the actual Ti content in the Zn bath. Thus, no Fe-Ti intermetallic compound will be formed at the Fe/Zn-Ti interface. This conclusion is correct despite the fact that the approximation of diluted solutions is not obeyed here (see the high mole fractions calculated in Table IX). This thermodynamic conclusion is supported by the results of the GD-OES measurements.

Now, let us briefly discuss the possible roles of the contaminants in Zn and Ti. Only elements having higher affinity to oxygen compared to Ti are considered. Thus, Pb, Cd, Fe, Cu, Sn, and Si of Tables II and III are excluded from this analysis, as they are not able to displace Ti from its oxide.[25] From the elements listed in Tables II and III, only Al and Mg can displace Ti from its oxide.[25] To analyze the maximum possible thickness of the $Al_2O_3$ and MgO layers, Eq. [4a] is applied accordingly. For a thin alumina layer, $d_{Al_2O_3,max}$ = 1.5 to 2.3 nm. Thus, between a thicker $TiO_2$ layer and a bulk Zn layer, a thin alumina layer is expected to form. This is confirmed by our SNMS measurements (see Figure 6). The calculated thickness of the thin magnesia layer is below 0.1 nm, i.e., MgO is not expected to form. Indeed, it is not found experimentally.

As a conclusion to this thermodynamic analysis, the Ti content of the Zn-Ti liquid alloy will not be segregated to the Fe/Zn-Ti interface, rather it will be segregated to the Zn-Ti/air interface, due to the formation of a thin $TiO_2$ layer. Due to Al contamination of the Ti shavings, a very thin $Al_2O_3$ sub-layer will be formed at the $TiO_2$/Zn interface. The thickness of the $TiO_2$ layer is limited by the amount of Ti in the Zn-Ti coating, at least, for long cooling times. The actual thickness of the $TiO_2$ layer is below this theoretical limit, due to the limited cooling time, while the Zn-Ti coating remains liquid on the surface of the steel sample after it is removed from the Zn-Ti bath. When the Zn-Ti layer solidifies and cools together with the steel sample, the diffusion of Ti atoms through the Zn layer and the diffusion of $Ti^{+4}$ ions through the $TiO_2$ layer are expected to slow down significantly. Thus, the thickness and the cooling conditions of the Zn-Ti coating will be essential to model the thickness of the $TiO_2$ layer. Therefore, these subjects are discussed in the next sub-chapters.

### B. Adhesion (the Thickness of the Zn layer)

The goal of this sub-chapter is to model the thickness of the Zn-Ti coating, adhered to the steel sample, denoted by $d_{Zn}$ (m). As the flow of liquid zinc along the vertical steel sheet was not observed during the cooling phase, this effect (flow and viscosity of liquid zinc) is not taken into account in this analysis.

For simplicity, the adhesion of pure liquid Zn will be modeled, as a small concentration of Ti is not expected to change the thickness of the Zn-Ti layer significantly. Thus, $d_{Zn}$ is taken as independent on the Ti content of the Zn bath. Let us consider a vertical, solid steel sheet with a thickness of $d$, width of $w$, and length of $L$. It is submerged into pure liquid Zn, and then it is slowly withdrawn from it in a vertical direction, along its length $L$. Suppose the steel sheet is perfectly wettable by the liquid Zn, and therefore, a thin Zn layer is adhered to it on both sides, with an average thickness of $d_{Zn}$. Using the principle outlined in,[34,35] the following equation can be derived for the interfacial force, pulling the adhered liquid layer upwards, along one side of the steel sheet:

$$F_{int} = w \times (\sigma_{sg} - \sigma_{sl} - \sigma_{lg}), \quad [6a]$$

where $\sigma_{sg}$ (J/m²) is the surface energy of the solid steel, $\sigma_{lg}$ (J/m²) is the surface tension of the liquid zinc, and $\sigma_{sl}$ (J/m²) is the interfacial energy between solid steel and liquid Zn. The steel sheet is kept vertically in air by the equipment. However, the liquid zinc layer hangs attached to it and is dragged by the gravity force down, according to the following equation:

$$F_{grav} = w \times L \times d_{Zn} \times \rho \times g, \quad [6b]$$

where $\rho = 6{,}580$ kg/m³[24] is the density of the liquid zinc, and $g = 9.81$ m/s² is the acceleration due to gravity during our experiments. Making equal the upward- and downward-oriented forces given in Eqs. [6a] and [6b], the thickness of the Zn layer can be expressed as

$$d_{Zn} = \frac{\sigma_{sg} - \sigma_{sl} - \sigma_{lg}}{L \times \rho \times g}. \quad [6c]$$

The height of the sample is about: $L = 0.1$ m (see Table V). The solid/liquid interfacial energy of the steel/zinc couple is estimated using a model[36]: $\sigma_{sl} \cong 0.3 \pm 0.1$ J/m². The major problem in applying Eq. [6c] is that during the removal of the sample and before solidification of the liquid Zn layer, both the upper edge of the steel sheet and liquid Zn are gradually oxidized, i.e., their surface energy and surface tension values gradually change. Therefore, two extreme cases will be considered here: one without and one with full oxidation.

When oxidation of both steel and Zn is neglected, the following values can be applied[24,36]: $\sigma_{sg} \cong 2.3 \pm 0.1$ J/m², $\sigma_{lg} \cong 0.8 \pm 0.1$ J/m². Substituting these and the above values into Eq. [6c], $d_{Zn} = 185 \pm 45$ μm results. When oxidation of both steel and Zn is allowed to go to full extent, the following values can be applied[37]: $\sigma_{sg} \cong 0.7 \pm 0.1$ J/m², $\sigma_{lg} \cong 0.6 \pm 0.1$ J/m² (note: the surface tensions of ZnO and $TiO_2$ are similar). Substituting these and the above values into Eq. [6c], $d_{Zn} = -30 \pm 45$ μm. Thus, during oxidation, the thickness of the adhered layer decreases considerably, from around 180 μm for the non-oxidized case to around zero (or 15 μm in the best case) for the fully oxidized case. This predicted interval reproduces the measured interval of 60 to 30 μm (see Table IV).

The theory presented here also predicts well the dependence of the thickness of the adhered Zn layer on the bath temperature. At low bath temperature (just above the melting point of liquid Zn), solidification time is very short (see the next sub-chapter), so both Fe and Zn are little oxidized, and therefore, the adhered Zn layer is relatively thick. On the other hand, at a higher bath temperature, solidification time is considerably increased, so the extent of oxidation of both Fe and Zn increases, and therefore, the thickness of the adhered Zn layer becomes relatively thin.

Using the theoretical Eq. [6c] for the L-dependence of $d_{Zn}$, and the experimental data for the T-dependence of $d_{Zn}$ (see Figure 7), the final semi-empirical equation follows as

$$d_{Zn} \cong \frac{117}{L \times T_{Zn}^{2.55}}, \quad [6d]$$

where L is substituted in m, $T_{Zn}$ is substituted in K, and $d_{Zn}$ results in m.

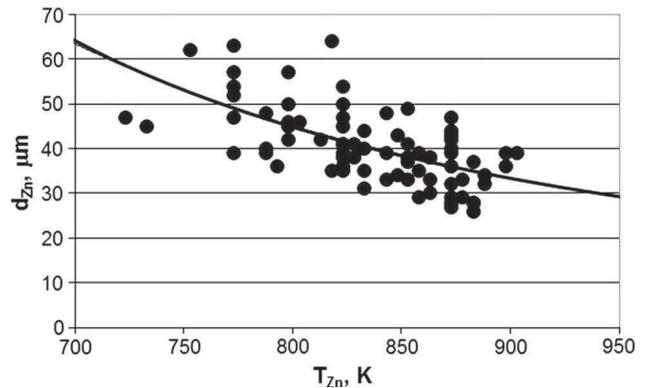

Fig. 7—The thickness of the Zn-Ti coating on the steel sample as a function of bath temperature. The solid dots are the values measured by GD-OES (see also Table IV). The line is calculated by Eq. [7d] and $L = 0.1$ m.

## C. Heat Flow (Cooling Time to Solidification)

The goal of this sub-chapter is to model the time period from the moment of removal of the steel sheet from the Zn-Ti bath (*i.e.*, from the moment when the Zn-Ti liquid alloy comes into contact with oxygen molecules of air) to the moment of starting the solidification of the Zn-Ti coating on the steel sheet (at this moment, the diffusional growth of the outer $TiO_2$ layer practically stops). In this simplified analysis, only the cooling of the steel sample is considered, as the thin Zn-Ti layer has a negligible influence on the cooling time.

The bath temperature, and therefore the sample temperature after its removal from the bath, is too low for radiation heat transfer. Therefore, cooling of the sample is modeled here by the spontaneous convection of air along the surface of the sample, caused by the temperature difference between the sample and outside still air. It is supposed that the sample cools according to the Newton law, *i.e.*, the temperature gradient across the thickness of the sample is negligible. This is the case if the following condition is fulfilled[30]:

$$\mathrm{Nu} \equiv \frac{h \times d}{k} < 0.1, \quad [7a]$$

where Nu (dimensionless) is the Nusselt number, $k$ (W/mK) is the heat conductivity of the steel sample (its value is about 50 W/mK[30]), and $h$ (W/m$^2$K) is the heat transfer coefficient at the steel/air interface. The latter is described approximately as[30]

$$h \cong 1.42 \times \left(\frac{\Delta T}{L}\right)^{1/4} \quad \text{if} \quad 5 \times 10^{-6} \mathrm{m}^3 \mathrm{K} < L^3 \\ \times \Delta T < 50 \, \mathrm{m}^3 \mathrm{K}, \quad [7b]$$

where $\Delta T \equiv T_{Zn} - T_{air}$ (K), *i.e.*, the difference between the temperatures of the liquid bath and that of air. In Table V, L = 0.1 m and $\Delta T$ = 400 K to 620 K (400 °C to 620 °C), and thus, $L^3 \times \Delta T$ = 0.40 to 0.62 m$^3$K. Therefore, Eq. [7b] is valid for our case. Substituting the average value of $\Delta T$ = 510 K (510 °C) into Eq. [7b], the average value of $h$ = 12.0 W/m$^2$K. Substituting this value, $k$ = 50 W/mK and $d$ = 8.10$^{-4}$ m (see Table V), into Eq. [7a], Nu = 1.9 × 10$^{-4}$. Comparing this value with the criterion of Eq. [7a], it follows that our case indeed obeys the Newton law of cooling.

Now let us make equal the rate of heat loss by the steel sample and the rate of heat gain by the spontaneous air flow:

$$-\frac{L \times w \times d}{V_{Fe}} \times C_p \times \frac{dT}{dt} = h \times 2 \times L \times w \times (T - T_{air}), \quad [7c]$$

where $V_{Fe}$ (m$^3$/mol) is the molar volume of steel, $C_p$ (J/molK) is the molar heat capacity of steel, $T$ (K) is the average temperature of steel, $t$ (s) is the cooling time from the moment of removal of the steel sample from the Zn bath, and $-dT/dt$ (K/s) is the cooling rate of the steel sample. Let us re-organize Eq. [7c] as

$$\frac{dT}{(T - T_{air})} = -\frac{2 \times h \times V_{Fe}}{d \times C_p} \times dt. \quad [7d]$$

To solve Eq. [7d], the following boundary condition is used: at $t = 0$ s, $T = T_{Zn}$. To simplify the integration of Eq. [7d], expression $h \times V_{Fe}/(d \times C_p)$ is supposed to have a constant value in the whole temperature interval of $T_{Zn}$ = 703 K to 923 K (430 °C to 650 °C). Then, the solution of Eq. [7d] is as follows:

$$T = T_{air} + (T_{Zn} - T_{air}) \times \exp\left(-\frac{2 \times h \times V_{Fe}}{d \times C_p} \times t\right). \quad [7e]$$

According to Eq. [7e], at $t = 0$, $T = T_{Zn}$, while at $t \to \infty$, $T \to T_{air}$. The goal is to calculate the cooling time, needed to reach the solidification temperature of pure Zn, $T_m$ = 693 K (420 °C). Therefore, let us substitute $T = T_m$ into Eq. [7e] and express from it t, to be called hereinafter the cooling time and to be denoted as $t_c$ (s):

$$t_c = \frac{d \times C_p}{2 \times h \times V_{Fe}} \ln\left(\frac{T_{Zn} - T_{air}}{T_m - T_{air}}\right). \quad [7f]$$

The required thermo-physical parameters of steel in the T interval of 703 K to 923 K (430 °C to 650 °C) are written as[23,25,38] (with $T$ in K)

$$C_p \cong 4.85 + 0.0425 \times T \quad \mathrm{J/molK}, \quad [7g]$$

$$V_{Fe} \cong 7.07 \times 10^{-6}(1 + 4.8 \times 10^{-5} \times (T - 273)) \quad \mathrm{m}^3/\mathrm{mol}. \quad [7h]$$

These thermo-physical quantities should be calculated at a temperature being the average of the bath temperature and the melting point of Zn. Thickness $d$ in Eq. [7f] is taken as the thickness of the steel sheet plus the double of the average Zn coating thickness, calculated by Eq. [6d]. The results of calculations are shown in Figure 8, using the parameters of Table V. The calculated data points are fitted by the following approximated equation ($T_{Zn}$ in K, $t_c$ in seconds):

$$t_c \cong -1920 + 293.6 \times \ln T_{Zn}. \quad [7i]$$

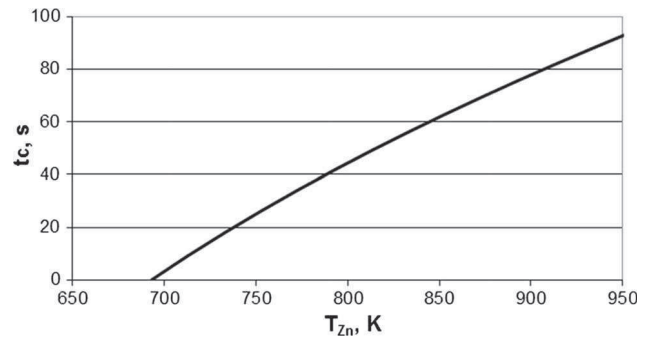

Fig. 8—The cooling time of the Zn-Ti coated steel sample from the moment of its removal from the Zn-Ti bath to the moment when its temperature reaches the melting point of Zn ($T_m$ = 693 K (420 °C)).

Equation [7i] is valid only for the parameter values of Table V and for the bath temperature interval of $T_{Zn}$ = 703 K to 923 K (430 °C to 650 °C). In this T interval, it takes 5 to 85 seconds for the steel sample together with its liquid Zn-Ti coating to cool from the bath temperature to the solidification temperature of Zn. This is the time period, during which the $TiO_2$ layer can grow on the outer surface of the Zn-Ti coating. The next sub-chapter is devoted to modeling its thickness.

### D. Reaction Kinetics and Diffusion (the Thickness of the $TiO_2$ Layer)

There is a steel sheet with a Zn-Ti coating of thickness $d_{Zn}$ (see Eq. [6d]), with a Ti content of $C_{Ti}$ = 0.15 wt pct (see Table V), which cools within a limited time period of time $t_c$ (see Eq. [7i]) from the temperature of the Zn-Ti bath to the solidification temperature of Zn. Only during this limited period of time the formation and thickening of the $TiO_2$ layer is possible on the outer surface of the Zn-Ti coating. The goal of this sub-chapter is to model its thickness. The kinetics of its formation and growth includes the following consecutive steps:

(i) the diffusion of Ti atoms through the liquid Zn layer to the liquid/gas surface or (after the thin $TiO_2$ layer is formed on it) to the liquid Zn/solid $TiO_2$ interface;
(ii) the ionization of Ti atoms into $Ti^{+4}$ ions at the liquid Zn/solid $TiO_2$ interface, and at the same time the dissociation of $O_2$ molecules and the ionization of two O atoms at the solid $TiO_2$/gas interface into two $O^{-2}$ ions. However, ionization will accumulate positive and negative charges at the two sides of the $TiO_2$ layer; thus, ionization would stop, if the separated charges (i.e., the ions, carrying those charges) are not brought to the proximity of each other; and
(iii) the latter is the driving force for the diffusion of the $Ti^{+4}$ ions from the liquid Zn/solid $TiO_2$ interface to the solid $TiO_2$/gas surface; when this ion transfer is completed, the neutralization of charges of two $O^{-2}$ ions by one $Ti^{+4}$ ion at the solid $TiO_2$/gas interface takes place; in this way, the $TiO_2$ ionic lattice is grown by one new "molecular" layer toward the gas phase.

In the above mechanism, it was supposed that the diffusion of $Ti^{+4}$ ions through $TiO_2$ is much faster than that of the $O^{-2}$ ions, which is in agreement with their different ionic radii (see above and also below). It was also supposed that the diffusion of the $O_2$ molecules from the bulk of the gas phase to the $TiO_2$/gas interface is not a rate-limiting step of $TiO_2$ growth, due to the high partial pressure of $O_2$ in air, and also due to the high diffusion coefficient of $O_2$ in air. Let us further suppose that among the above processes, ionization is much probably not a rate-limiting step of $TiO_2$ growth. Thus, only two processes can be the rate-limiting steps of $TiO_2$ growth: the diffusion of Ti atoms through the Zn bath, or the diffusion of the $Ti^{+4}$ ions through the $TiO_2$ layer. These two processes will be considered in the next sub-chapters.

#### 1. Diffusion of Ti atoms through the liquid Zn bath

As was shown in sub-chapter IV-A, the Ti atoms of the Zn-Ti liquid alloy and the $O_2$ molecules of air react to form a $TiO_2$ compound, and as a result, the Ti content of the Zn bath decreases by more than 10 orders of magnitude in equilibrium (see Table VII), i.e., at the liquid Zn/$TiO_2$ interface. This concentration reduction will create a Ti-concentration gradient from the bulk of the Zn-Ti bath toward to Zn/$TiO_2$ interface. Suppose that the Ti atoms will "disappear" (i.e., transform into $TiO_2$) immediately as they arrive to the Zn/$TiO_2$ interface, i.e., suppose that diffusion of Ti through the Zn bath is the rate-limiting step of $TiO_2$ growth. Then, one can write the equation for the ratio of the average Ti content of the Zn bath after some diffusion time t, to the initial average Ti content of the Zn bath[30]:

$$\frac{C_{Ti,t}}{C_{Ti}} \cong \exp\left(-\frac{D_{Ti} \times \pi^2}{4 \times d_{Zn}^2} \times t\right), \quad [8]$$

where $D_{Ti}$ (m²/s) is the diffusion coefficient of Ti atoms through liquid Zn, $C_{Ti}$ (wt pct) is the initial Ti content of the Zn bath (= 0.15 wt pct, see Table V), $C_{Ti,t}$ is the average Ti content in the Zn bath after period of time $t$ (s). The order of magnitude of the diffusion coefficients through liquid metals is $10^{-9}$ m²/s.[24,40] Let us consider the average thickness of the Zn-Ti layer of $d_{Zn}$ = 45 μm (see Table IV) and the average cooling / diffusion time of $t_c$ = 40 seconds (see Figure 7). Substituting these values into Eq. [8], the result is $C_{Ti,40s}/C_{Ti}$ = 7 × $10^{-22}$. Thus, under the average conditions of our experiments, practically all the initial Ti content of the Zn-Ti layer is able to diffuse to the Zn / $TiO_2$ interface, if it is supposed that those atoms are immediately built into the $TiO_2$ layer. If this was the case, Eq. [4b] would determine the thickness of the $TiO_2$ layer. For the value of $x_{Ti}$ = 2 × $10^{-3}$ (see Table V) and $d_{Zn}$ = 45 μm (see above and Table IV), this would lead to the thickness of the $TiO_2$ layer of 186 nm. However, this value is much higher than the values obtained in our experiments (see Table IV). This contradiction means that the above hypothesis is wrong, i.e., in fact the diffusion of Ti atoms through the Zn bath is not the rate-limiting step of $TiO_2$ growth. As follows from the mechanism given above, the rate-limiting step of $TiO_2$ growth can only be the diffusion of $Ti^{+4}$ ions through the ionic lattice of $TiO_2$. The next sub-chapter is devoted to this question.

Before going on, one should mention that when Zn is solidified, the diffusion coefficient of Ti through solid Zn would immediately drop by at least 4 orders of magnitude.[30] Keeping all other parameters constant, that would lead to $C_{Ti,40s}/C_{Ti}$ = 0.995 instead of 7 × $10^{-22}$ (see above). This would provide only 0.5 pct of the total Ti content of the Zn-Ti coating to be converted into $TiO_2$. According to Eq. [4b], this would mean only $d_{TiO_2}$ = 1 nm, which is practically nil. It means that when the Zn-Ti coating is solidified, the diffusion of Ti through it would immediately stop, meaning the end of the growth of the $TiO_2$ layer.

When the Zn-Ti layer is cooled to the room temperature, the diffusion coefficient of Ti through solid Zn would decrease by about further 6 orders of magnitude. This means that to grow a 10-nm-thick $TiO_2$ layer at room temperature (supposing the diffusion of Ti through solid Zn is the rate-limiting step), the time period of $4.2 \times 10^9$ seconds, or 133 years is needed. This analysis shows why the time period of $TiO_2$ growth was limited to the cooling time of the Zn-Ti coating from the bath temperature to the solidification temperature ($t_c$). This analysis also shows that the color of the hot-dip galvanized steel sheet will be stable for about 100 years from the moment of its production.

2. *Diffusion of $Ti^{+4}$ ions through the ionic lattice of $TiO_2$*

It was proven above that the rate-limiting step of $TiO_2$ growth is the diffusion of $Ti^{+4}$ ions through the ionic lattice of $TiO_2$. Then, the thickness of the $TiO_2$ layer is written as[30]

$$d_{TiO_2} = \sqrt{2 \times D_{Ti^{+4}} \times C_{Ti^{+4}} \times V_{TiO_2} \times t_c}, \quad [9a]$$

where $D_{Ti^{+4}}$ (m$^2$/s) is the diffusion coefficient of $Ti^{+4}$ ions through $TiO_2$, $V_{TiO_2} = 1.9 \cdot 10^{-5}$ m$^3$/mol[31] is the molar volume of $TiO_2$, $t_c$ (s) is the cooling time, and $C_{Ti^{+4}}$ (mol/m$^3$) is the molarity of the $Ti^{+4}$ ions at the liquid Zn/$TiO_2$ interface. As the $TiO_2$ layer grows, less and less Ti atoms will remain in the Zn bath, and so the value of $C_{Ti^{+4}}$ will be also gradually decreasing. The average value of $C_{Ti^{+4}}$ to be used in this model can be calculated using the following equation:

$$C_{Ti^{+4}} = 1.0 \times 10^5 \left( x_{Ti} - 0.242 \times \frac{d_{TiO_2}}{d_{Zn}} \right), \quad [9b]$$

where parameter $1.0 \times 10^5$ mol/m$^3$ is the reciprocal of the molar volume of pure Zn at its melting point, the term in the brackets is the time average mole fraction of the Ti content in the liquid Zn-Ti alloy, and the second term in the brackets is half of the decrease of the Ti mole fraction in the Zn bath due to the formation of a $TiO_2$ layer of thickness $d_{TiO_2}$, expressed from Eq. [4b].

Let us consider the example of average blue sample in Table IV, obtained at about $T_{Zn} = 873$ K (600 °C) with $d_{Zn} = 35$ μm and $d_{TiO_2} = 69$ nm at $x_{Ti} = 2.0 \times 10^{-3}$ and substitute these values into Eq. [9b]: $C_{Ti^{+4}} = 152$ mol/m$^3$. The corresponding cooling time is $t_c = 69$ seconds, as follows from Eq. [7i]. Unfortunately, the diffusion coefficient of $Ti^{+4}$ ions through $TiO_2$ is not known. So, let us estimate its value from Eq. [9a], and using the above values, $D_{Ti^{+4}} = 1.19 \times 10^{-14}$ m$^2$/s. This value corresponds to a temperature somewhere between 873 K (600 °C = the bath temperature) and 693 K (420 °C) is the solidification temperature of Zn. As the diffusion coefficient is an exponential function of temperature, the average temperature is found as a combination of about 70 pct of the higher T and about 30 pct of the lower T:

$$T_{av} \cong 208 + 0.7 \times T_{Zn}, \quad [9c]$$

where parameter 208 K = $0.3 \times 693$ K. Substituting $T_{Zn} = 873$ K (600 °C) into Eq. [9c], $T_{av} = 819$ K (546 °C).

For comparison, the diffusion coefficient of $O^{-2}$ ions through the ionic lattice of solid $TiO_2$ is in the interval of $D_{O^{-2}} = 10^{-22}$ to $10^{-24}$ m$^2$/s, at a temperature around $T_{av} = 819$ K (546 °C).[31] One can see that this is lower by at least 8 orders of magnitude compared to the diffusion of $Ti^{+4}$ ions through the same lattice, found above ($1.19 \times 10^{-14}$ m$^2$/s). This comparison proves the validity of our hypothesis above that the growth of $TiO_2$ takes place along its outside plane, thanks to the much faster diffusion of $Ti^{+4}$ ions through this layer, *i.e.*, the role of $O^{-2}$ diffusion in $TiO_2$ growth is negligible.

Now, let us consider another example of Table IV: the average yellow sample, obtained at about $T_{Zn} = 814$ K (541 °C) with $d_{Zn} = 43$ μm and $d_{TiO_2} = 36$ nm at $x_{Ti} = 2.0 \times 10^{-3}$. Substituting these values into Eq. [9b], the molarity of the $Ti^{+4}$ ions is $C_{Ti^{+4}} = 180$ mol/m$^3$. The corresponding cooling time is $t_c = 49.4$ seconds, as follows from Eq. [7i]. The estimated value for the diffusion coefficient of $Ti^{+4}$ ions through $TiO_2$ from Eq. [9a] is $D_{Ti^{+4}} = 3.84 \times 10^{-15}$ m$^2$/s. This value corresponds to $T_{av} = 778$ K (505 °C), as follows from Eq. [9c]. Using these two values of $D_{Ti^{+4}}$ estimated at two different $T_{av}$ values, its T-dependence is written approximately as (temperature is measured in K):

$$D_{Ti^{+4}} \cong 2.9 \times 10^{-5} \times \exp\left(\frac{-17,700}{T_{av}}\right). \quad [9d]$$

It is worth to mention that both the pre-exponential coefficient ($2.9 \times 10^{-5}$ m$^2$/s) and the activation energy (147 kJ/mol) of Eq. [9d] fit other literature data on the diffusion of metallic ions through oxide lattices,[31] so Eq. [9d] seems to be reasonable. Now, let us substitute Eq. [9b] into Eq. [9a], and let us express from here the final equation for the required quantity, $d_{TiO_2}$:

$$d_{TiO_2} = 0.5 \times \left(-k_2 + \sqrt{k_2^2 + 4 \times k_1}\right), \quad [9e]$$

$$k_1 = 2 \times 10^5 \times D_{Ti^{+4}} \times V_{TiO_2} \times t_c \times x_{Ti}, \quad [9f]$$

$$k_2 = 4.84 \times 10^4 \times D_{Ti^{+4}} \times V_{TiO_2} \times \frac{t_c}{d_{Zn}}. \quad [9g]$$

Using Eqs. [9c] through [9g], [6d] and [7i] and the values of $V_{TiO_2} = 1.9 \times 10^{-5}$ m$^3$/mol and $x_{Ti} = 0.002$ (see Table V), the thickness of the $TiO_2$ layer is calculated as function of bath temperature for the conditions of our experiments (see Figure 9). As it can be seen from Figure 9, the calculated line reproduces perfectly the three average measured values (see Table IV). However, it should be reminded that two of these three points were used to estimate the diffusion coefficient of $Ti^{+4}$ through solid $TiO_2$, so this agreement is surprising only for samples with violet color. At this point, it should also be mentioned that only the two numerical parameters of Eq. [9d] were found by fitting experimental data in this paper; no other fitting parameters are used here.

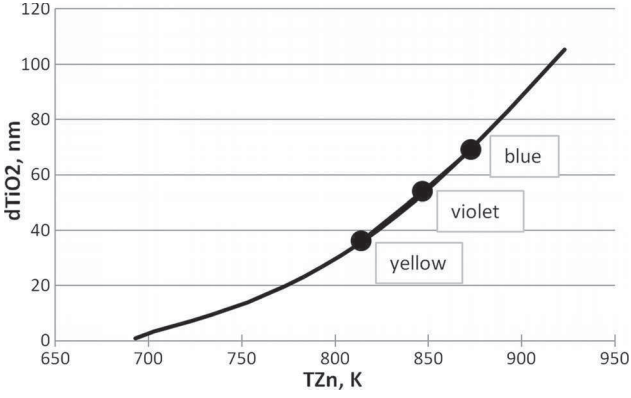

Fig. 9—The thickness of the TiO$_2$ layer, as function of the bath temperature. Data points are from the experiments (Table IV). The line is calculated by Eqs. [9c] through [9g], [6d], [7i] for the conditions of our experiments (Table V, $V_{\text{TiO}_2} = 1.9 \times 10^{-5}$ m$^3$/mol[31]).

The validity of Eqs. [9e] through [9g] is limited at $T_{\text{Zn}} \geq 693$ K (420 °C), which is the melting point of Zn. However, the above equations also have an upper limiting temperature, dictated by the combination of Eqs. [4b] and [6d]:

$$d_{\text{TiO}_2,\max} = \frac{242 \times x_{\text{Ti}}}{L \times T_{\text{Zn}}^{2.55}}. \qquad [9h]$$

Comparing the results obtained from Eq. [9h] and Eqs. [9c] through [9g], [6d] and [7i] for parameters of Table V, they become equal at $T_{\text{Zn}} = 946$ K (673 °C) corresponding to $d_{\text{TiO}_2} = 125$ nm. This is the maximum TiO$_2$ thickness, which can be obtained under our fixed experimental conditions. Thus, the validity of Eqs. [9e] through [9g] is limited at 693 K (420 °C) $\leq T_{\text{Zn}} \leq$ 946 K (673 °C).

3. *The effect of Al contamination*

As it is shown in Section IV–A, a maximum of 1.5- to 2.3-nm-thick alumina layer can be formed with the oxygen of air, if all the Al content of the Zn bath is consumed. As aluminum has a higher affinity to oxygen compared to titanium, first this 1.5- to 2.3-nm-thin alumina layer will form on the top of the Zn bath. The diffusion coefficient of Ti$^{+4}$ ions through this thin alumina layer is probably similar to the one through the TiO$_2$ layer. Thus, an outer TiO$_2$ layer will be growing toward the air phase, as described above. The only correction is that the thickness values calculated above will correspond to the total thickness of the Al$_2$O$_3$-TiO$_2$ double surface oxide layer, with only 1.5 to 2.3 nm Al$_2$O$_3$ positioned below the thicker TiO$_2$ layer.

E. *Optics (the Final Color of the Steel Sheet)*

In this sub-chapter, the thickness of the outer TiO$_2$ layer is connected to the color of the steel/Zn-Ti/TiO$_2$ sample (for simplicity, the effect of a much thinner alumina layer is neglected). Titanium dioxide is a transparent dielectric. An incident light is partly reflected from its surface and partly transmitted through. The transmitted part of the light reflects fully back from the metallic Zn layer underneath. A part of this reflected light is transmitted through the upper surface of the oxide film and interferes with the originally reflected light. As a result, both constructive and destructive interference will take place at special wavelengths. For a light, being normal to the surface and for the case when TiO$_2$ is on the top of a higher refractive index material (such as Zn), these special wavelengths are written as[41,42]

$$\lambda_{\text{dest}} = \frac{2 \times n_{\text{TiO}_2} \times d_{\text{TiO}_2}}{m - 0.5}, \qquad [10a]$$

$$\lambda_{\text{const}} = \frac{2 \times n_{\text{TiO}_2} \times d_{\text{TiO}_2}}{m}, \qquad [10b]$$

where $\lambda_{\text{des}}$ (m) is the special wavelength of destructive interference, $\lambda_{\text{const}}$ (m) is the special wavelength of constructive interference, $d_{\text{TiO}_2}$ (m) is the thickness of the TiO$_2$ layer with two parallel Zn/TiO$_2$ and TiO$_2$/air interfaces, and $n_{\text{TiO}_2}$ (dimensionless) is the refractive index of TiO$_2$, $m = 1, 2, \ldots$ (dimensionless) is the order of interference. The following approximated equation describes the wavelength dependent average refractive index of TiO$_2$[41]:

$$n_{\text{TiO}_2} \cong \frac{0.0101}{\lambda^{0.390}}. \qquad [10c]$$

Let us substitute Eq. [10c] into Eqs. [10a] and [10b] and express from them the special wavelengths of destructive and constructive interference:

$$\lambda_{\text{dest}} = \left(\frac{0.0202 \times d_{\text{TiO}_2}}{m - 0.5}\right)^{0.719}, \qquad [10d]$$

$$\lambda_{\text{const}} = \left(\frac{0.0202 \times d_{\text{TiO}_2}}{m}\right)^{0.719}. \qquad [10e]$$

To connect the color of the samples with the thickness of the TiO$_2$ layer, two more tables are needed in addition to Eqs. [10d] and [10e]. In Table X, the wavelength intervals, corresponding to different colors are collected. In Table XI, the color pairs are collected, showing color B appearing when color A is destructed, and *vice versa*. Based on Eqs. [10d] and [10e] and Tables X and XI, Table XII is constructed. In Table XII, the intervals of destructed and constructed wavelengths are given as function of the TiO$_2$ thickness, and finally the resulting color of the sample is given. For example, as follows from the second row of Table XII, when the thickness of the TiO$_2$ layer is below 32 nm, all destructive and constructive wavelengths calculated from Eqs. [10d] and [10e] will be below 400 nm (the lower limit of visible light by humans), and therefore, according to Table X, this interference will not disturb color perception by humans, *i.e.*, the color of the surface remains metallic. However, at larger values of $TiO_2$ thickness, interference will take place in the visible interval of wavelengths, and thus, the color will change, as shown in the last column of Table XII.

F. *Summary of Discussion*

The $TiO_2$-thickness values of Table XII can be connected with the bath temperatures using Figure 9 or Eqs. [9c] through [9g], [6d] and [7i]. In this way, finally our goal is achieved: bath temperature is connected to the color of the sample (see Table XIII). In Table XIII, the theoretical and experimental bath temperature intervals are compared, providing different colors. As follows from Table XIII, there is a generally good agreement between theory and experiments. The only missing experimental color is green, which is expected to appear theoretically. Let us mention that green color was not found in similar experiments of Le and Cui,[20] either. Further research is needed to explain the reasons of this controversy. One of the possible reasons is the formation of 1.5- to 2.3-nm-thin $Al_2O_3$ sub-layer at the interface of the outer $TiO_2$ layer and the macroscopic Zn layer (predicted theoretically in Section IV–A and proven experimentally in Figure 6).

Since the model reproduced reasonably well the experimental results as function of bath temperature, let us apply the same model to predict the role of different experimental parameters on the color of the sample:

– the color is expected to shift from metallic through yellow and violet toward blue by increasing the bath temperature (proven experimentally), the Ti content of the Zn bath, the thickness of the steel sheet, the air temperature, and the length of the sample;
– the color is expected to shift from blue through violet and yellow toward metallic by increasing the velocity of the cooling air (as it decreases the cooling time); and
– the color is expected to be independent on the width of the steel sample, on the velocity of pushing the sample into liquid Zn-Ti bath, on the holding time of the sample within the bath, on the removal velocity of the sample from the bath, on air pressure, and on the partial pressure of oxygen.

Table X. The Wavelength Intervals of Colors as Seen by Average Human Eye[41]

| Color | $\lambda$ (nm) |
|---|---|
| Violet | 400–450 |
| Blue | 450–490 |
| Green | 490–560 |
| Yellow | 560–590 |
| Orange | 590–635 |
| Red | 635–700 |

Table XI. Color Pairs (Color B Appears, When Color A Is Destructed and *Vice Versa*)[41]

| Color A | Color B |
|---|---|
| Blue | yellow |
| Violet | green |
| Red | turquoise |
| Orange | blue / turquoise |

Table XII. Destructed and Constructed Wavelengths and Colors Appearing as Function of $TiO_2$ Thickness (Calculated by Eqs. [10d] and [10e] and Tables X and XI)

| $d_{TiO_2}$ (nm) | $\lambda_{dest}$ (nm) (color) | $\lambda_{const}$ (nm) (color) | Final Color |
|---|---|---|---|
| ≤32 | ≤400 (not visible) | ≤400 (not visible) | remains metallic |
| 32–37 | 400–450 (violet) | ≤400 (not visible) | green |
| 37–42 | 450–490 (blue) | ≤400 (not visible) | yellow |
| 42–51 | 490–560 (green) | ≤400 (not visible) | violet |
| 51–54 | 560–590 (yellow) | ≤400 (not visible) | blue |
| 54–60 | 590–635 (orange) | ≤400 (not visible) | blue/turquoise |
| 60–64 | 635–670 (red) | ≤400 (not visible) | turquoise |
| 64–69 | 670–700 (red) | 400–430 (violet) | turquoise + violet |
| 69–75 | ≥700 (not visible) | 430–450 (violet) | violet |
| 75–84 | ≥700 (not visible) | 450–490 (blue) | blue |

Table XIII. Bath Temperature Intervals, Corresponding to Different Colors ($T_{Zn}$, K (°C)), According to Theory (from Table XII, Figure 9) and from the Experiments (Table IV)

| Color | Theory | Experiments | Evaluation |
|---|---|---|---|
| Metallic/silver | below 806 (533) | below 792 (519) | OK |
| Green | 806–817 (533–544) | — | ?? |
| Yellow | 817–827 (544–554) | 792–836 (519–563) | OK |
| Violet | 827–843 (554–570) | 837–857 (564–584) | OK |
| Blue/turquoise/violet | 843–894 (570–621) | 858–888 (585–615) | OK |

Let us specify that the above conclusions are valid within relatively small changes of the experimental parameters. For example, the color will remain metallic (*i.e.*, will be changed from other colors in contradiction to the above conclusions), if the oxygen pressure is decreased below a certain limit (as $TiO_2$ will not be able to form). Also, the color will remain metallic, if the Ti content of the Zn bath will be decreased below a certain limit. As shown in Table XII, if $d_{TiO_2} \leq 32$ nm, it will not cause any interference of the visible light. If this quantity is substituted into Eq. [9h], the lower limit of Ti is obtained, below which no color will be formed (L is in m, $T_{Zn}$ is in °C):

$$x_{Ti,min} = 1.3 \times 10^{-10} \times L \times T_{Zn}^{2.55}. \quad [11]$$

For example, for $L = 0.1$ m (same as in our experiments) and at $T_{Zn} = 873$ K (600 °C) the color will remain metallic, if $x_{Ti} \leq 4.1 \times 10^{-4}$, or if $C_{Ti} \leq 0.031$ wt pct. In reality, at $T_{Zn} = 873$ K (600 °C), a five times higher Ti content (0.15 wt pct) was applied than this theoretical limit for no color.

## V. CONCLUSIONS

1. The color of hot-dip galvanized steel was adjusted in a reproducible way using a liquid Zn-Ti metallic bath in air atmosphere and applying the bath temperature as the only adjustable experimental parameter. Coloring was found only for samples cooled in oxygen-containing gas (air) and for samples dipped into Ti-containing Zn bath. For steel samples of 0.8 mm thick and 100 mm long, dipped into a 0.15 wt pct Ti-containing Zn bath, the color remained metallic below 792 K (519 °C) of bath temperature; it was yellow at 814 K ± 22 K (541 °C ± 22 °C), violet at 848 K ± 10 K (574 °C ± 10 °C), and blue at 873 K ± 15 K (600 °C ± 15 °C).
2. The measured color parameter $\Delta b$ correlated well with the bath temperature.
3. Increasing the bath temperature, the thickness of the adhered Zn-Ti layer gradually decreased from 52 to 32 microns, as measured by GD-OES. It was probably due to the higher extent of oxidation of both steel and Zn-Ti alloy at higher bath temperature, which weakened the adhesion bond between solid Fe and liquid Zn-Ti.
4. The outer 100 nm of the Zn-Ti coating was found to be enriched in Ti, as measured by GD-OES. Using SNMS it was proven that the outer layer is $TiO_2$; its thickness was found to increase with the bath temperature from 24 to 69 nm. Thus, it is experimentally proven that the color is due to the formation of this thin $TiO_2$ layer outside of the Zn-Ti coating on the steel sample; different colors appear depending on the thickness of this layer. At the interface between the outer $TiO_2$ layer and the inner macroscopic Zn layer a thin (about 2 ± 1 nm) Al enrichment is found, which probably corresponds to $Al_2O_3$.
5. A complex model was built to connect the bath temperature and other experimental parameters with the color of the sample. For this purpose, known relationships of chemical thermodynamics, adhesion, heat flow, kinetics of chemical reactions, diffusion, and optics were applied. This complex model was able to reproduce reasonably well our experimental observations and allowed making new predictions. The latter can be of help for further research and development of colored hot-dip galvanized steel sheets.
6. By chemical thermodynamics it was shown that Ti is not interface active and not reactive at the Fe/Zn-Ti interface (this is experimentally confirmed by GD-OES). It was also found that $TiO_2$ will be the only stable oxide at the Zn-Ti / air interface if the Ti content of the Zn bath is above $10^{-10}$ wt pct (this is experimentally confirmed by SNMS for 0.15 wt pct Ti). It was also shown that a 1.5- to 2.3-nm-thin alumina layer can be formed at the interface between the outer $TiO_2$ layer and the macroscopic inner Zn layer (confirmed by SNMS).
7. By using the principles of optics, it is shown that if the thickness of the $TiO_2$ layer is below 32 nm, light interference takes place only in the invisible spectrum for humans (below 400 nm of wavelength), and thus, the metallic color will not change (at least, as seen by human eye). However, with the increasing thickness of $TiO_2$ above 32 nm, different colors will appear mainly due to destructive and partly due to constructive light interference. The thickness of $TiO_2$ layer was found to depend on the thickness of the Zn-Ti coating, on the Ti content of the Zn bath, on the thickness and length of the steel sample, and mainly on temperature of the Zn-Ti bath. All these parameters were connected with the color of the coating. Also, the minimum Ti content of the Zn bath was found, needed for coloring the coating (see Eq. [11]). For the length of the steel sample of 0.1 m, and bath temperature of 873 K (600 °C), coloring is predicted to be impossible below the Ti content of the Zn bath of 0.031 wt pct.

## ACKNOWLEDGMENTS

The corresponding author acknowledges the financial support of project K101781 of the Hungarian Academy of Sciences. The present results are achieved in the Center of Applied Materials Science and Nano-Technology at the University of Miskolc and within the TÁMOP-4.2.2.A-11/1/KONV-2012-0019 project, and carried out as part of the TÁ-MOP-4.2.2.D-15/1/KONV-2015-0017 project in the framework of the New Széchenyi Plan. The realization of this project is supported by the European Union, and co-financed by the European Social Fund. The authors thank Szendrő Galva Ltd for providing us with SHG Zn and to Salgó Metall Ltd for providing us with titanium shavings. They also thank Dr. Olivér Bánhidi for his help in ICP analysis of the Ti shavings and of the Zn-Ti alloy. The authors appreciate the help of color measurements provided by Dr. Tamás Szabó. The inspiration received from late Dr. Éva Dénes of Dunaferr Ltd (Hungary) is mostly appreciated at the beginning of this project.

## APPENDIX THERMODYNAMIC PROPERTIES OF THE Fe/Zn-Ti/O$_2$ System

Standard formation Gibbs energies of possible Zn oxides and Ti oxides are collected in Table AI.[25] The oxides are positioned along their increasing Ti content in Table AI.

As the Ti content in the Zn bath is below 1 mol pct, the Zn bath is modeled as an infinitely diluted solution in Ti. Thus, the activity coefficient of Ti ($\gamma_{Ti}^\infty$) is independent of its mole fraction:

$$R \times T \times \ln \gamma_{Ti}^\infty \equiv \Delta G_{Ti}^{E\infty}, \quad [A1]$$

where $\Delta G_{Ti}^{E\infty}$ (J/mol) is the partial molar excess Gibbs energy of Ti in liquid Zn, $R = 8.3145$ J/molK, the universal gas constant, and $T$ (K) is the absolute temperature. Unfortunately, there are no measured data available for $\Delta G_{Ti}^{E\infty}$. However, the sign of this quantity is expected as $\Delta G_{Ti}^{E\infty} < 0$, as follows from the existence of stable intermetallic compounds in the Zn-Ti binary equilibrium phase diagram.[26] From the Miedema model,[27] $\Delta H_{Ti}^\infty \cong -61$ kJ/mol, which is in agreement with the expected sign of $\Delta G_{Ti}^{E\infty}$. Combining this value with the most probable value of the excess entropy,[28] the most probable equation for the partial excess Gibbs energy of Ti is written as (in kJ/mol)

$$\Delta G_{Ti}^{E\infty} \cong -61 \times \exp(-T/3000). \quad [A2]$$

Substituting Eq. [A2] into Eq. [A1],

$$\gamma_{Ti}^\infty \cong \exp\left[\frac{-7340}{T} \times \exp\left(-\frac{T}{3000}\right)\right]. \quad [A3]$$

The values, calculated by Eq. [A3], are collected in Table AII at different temperatures. One can see that $\gamma_{Ti}^\infty < 1$, and also that with the increasing temperature, its value tends toward 1, i.e., the real solution tends toward the ideal solution, in accordance with the general law.[29]

Now, let us estimate the standard Gibbs energies of formation of intermetallic compounds FeZn and FeTi (in kJ/mol):

$$\Delta_f G_{FeZn}^\circ \cong -8.0 \times \left(1 - \frac{T}{\tau}\right), \quad [A4]$$

$$\Delta_f G_{FeTi}^\circ \cong -62 \times \left(1 - \frac{T}{\tau}\right). \quad [A5]$$

The numerical values (−8.0 and −62 kJ/mol) in Eqs. [A4] and [A5] are the standard heats of formation of the intermetallic compounds,[32] while parameter $\tau \cong$

Table AI. Standard Formation Gibbs Energies of Oxides ($\Delta_f G_i^\circ$, kJ/mol)[25]*

| T [K (°C)] | ZnO | Zn$_2$TiO$_4$ | TiO$_2$ | Ti$_4$O$_7$ | Ti$_3$O$_5$ | Ti$_2$O$_3$ | TiO |
|---|---|---|---|---|---|---|---|
| 700 (427) | −280.6 | −1388.0 | −815.9 | −2960.2 | −2137.7 | −1320.6 | −474.3 |
| 800 (527) | −269.7 | −1349.3 | −798.0 | −2899.0 | −2095.2 | −1293.8 | −464.9 |
| 900 (627) | −258.9 | −1310.9 | −780.2 | −2838.5 | −2053.0 | −1267.4 | −455.5 |
| 1000 (727) | −248.2 | −1272.8 | −762.5 | −2778.5 | −2011.0 | −1241.3 | −446.3 |
| $x_{Ti(oxide)}$ | 0 | 0.143 | 0.333 | 0.364 | 0.375 | 0.400 | 0.500 |

*Although values are given with the accuracy of 0.001 kJ/mol in,[25] to our opinion only the accuracy of 0.1 is justified.

Table AII. Activity Coefficients of Ti in the Diluted Solution of Liquid Zn at Different Temperatures Estimated by Eq. [A3]

| T [K (°C)] | 700 (427) | 800 (527) | 900 (627) | 1000 (727) |
|---|---|---|---|---|
| $\gamma_{Ti}^\infty$ | $2.5 \times 10^{-4}$ | $8.9 \times 10^{-4}$ | $2.4 \times 10^{-3}$ | $5.2 \times 10^{-3}$ |

Table AIII. Estimated Gibbs energies of formation of FeTi and FeZn intermetallic compounds, calculated by Eqs. [A4] and [A5]

| T [K (°C)] | 700 (427) | 800 (527) | 900 (627) | 1000 (727) |
|---|---|---|---|---|
| $\Delta_f G_{FeZn}^\circ$, kJ/mol | −7.3 | −7.2 | −7.1 | −7.0 |
| $\Delta_f G_{FeTi}^\circ$, kJ/mol | −56.8 | −56.0 | −55.3 | −54.5 |

8300 K (8027 °C) is the average value for the intermetallic compounds.[33] The values, calculated by Eqs. [A4] through [A5], are given in Table AIII.